\documentclass{article}
\pdfoutput=1

     \usepackage[preprint]{neurips_2023}

\usepackage[utf8]{inputenc} 
\usepackage[T1]{fontenc}    
\usepackage{comment}
\usepackage{url}            
\usepackage{booktabs}       
\usepackage{amsfonts}       
\usepackage{nicefrac}       
\usepackage{microtype}      
\usepackage[svgnames]{xcolor}
\usepackage{smile}
\usepackage{enumerate,natbib,mathrsfs}
\usepackage{algorithm}
\usepackage{algorithmic}
\usepackage{tablefootnote}
\usepackage{extarrows}
\usepackage[OT1]{fontenc}
\usepackage[bookmarks=false]{hyperref}

\hypersetup{
  pdftex,
  pdffitwindow=true,
  pdfstartview={FitH},
  pdfnewwindow=true,
  colorlinks,
  linktocpage=true,
  linkcolor=Red,
  urlcolor=Red,
  citecolor=Blue
}
\usepackage{stmaryrd}
\usepackage{graphicx}

\usepackage{multirow,multicol,arydshln} 
\usepackage{wrapfig}

\renewcommand{\d}[1]{\operatorname{d}\!#1}
\newcommand{\STTS}{\texttt{STTS}}
\newcommand{\VTTS}{\texttt{VTTS}}
\newcommand{\BR}{\texttt{BR}}
\newcommand{\Random}{\texttt{Random}}
\newcommand{\BBTS}{\texttt{BBTS}}

\title{Sequential Best-Arm Identification with Application to Brain-Computer Interface}

\author{%
  Xin Zhou \\
  Department of Biostatistics and Epidemiology\\
  University of California at Berkeley\\
  \texttt{xinzhou@berkeley.edu} \\
   \And
   Botao Hao \\
   Google Deepmind \\
   \texttt{haobotao000@gmail.com} \\
   \AND
   Jian Kang \\
   Department of Biostatistics \\
   University of Michigan \\
   \texttt{jiankang@umich.edu} \\
  \And
   Tor Lattimore \\
   Google Deepmind \\
   \texttt{lattimore@google.com } \\
   \And
   Lexin Li\thanks{Corresponding to Botao Hao, Lexin Li.} \\
Department of Biostatistics and Epidemiology\\
  University of California at Berkeley\\
   \texttt{lexinli@berkeley.edu} \\
}

\begin{document}

\maketitle

\begin{abstract}
A brain-computer interface (BCI) is a technology that enables direct communication between the brain and an external device or computer system. It allows individuals to interact with the device using only their thoughts, and holds immense potential for a wide range of applications in medicine, rehabilitation, and human augmentation. An electroencephalogram (EEG) and event-related potential (ERP)-based speller system is a type of BCI that allows users to spell words without using a physical keyboard, but instead by recording and interpreting brain signals under different stimulus presentation paradigms. Conventional non-adaptive paradigms treat each word selection independently, leading to a lengthy learning process. To improve the sampling efficiency, we cast the problem as a sequence of best-arm identification tasks in multi-armed bandits. Leveraging pre-trained large language models (LLMs), we utilize the prior knowledge learned from previous tasks to inform and facilitate subsequent tasks. To do so in a coherent way, we propose a sequential top-two Thompson sampling (\STTS) algorithm under the fixed-confidence setting and the fixed-budget setting. We study the theoretical property of the proposed algorithm, and demonstrate its substantial empirical improvement through both synthetic data analysis as well as a P300 BCI speller simulator example. 
\end{abstract}

\section{Introduction}

A brain-computer interface (BCI) is a groundbreaking technology that enables direct communication between the brain and an external device or computer system. It involves the use of various sensors, such as electroencephalography (EEG), electrocorticography (ECoG), or implantable neural electrodes, which detect and record the electrical signals produced by the brain. Those signals are then processed by machine learning algorithms to interpret and extract meaningful commands and intentions. BCI holds immense potential for a wide range of applications. For instance, it provides a valuable communication aid for individuals with disabilities \citep{wolpaw2018independent}. 

\begin{figure}[t!]
\centering
\includegraphics[width=8.75cm,height=3.6cm]{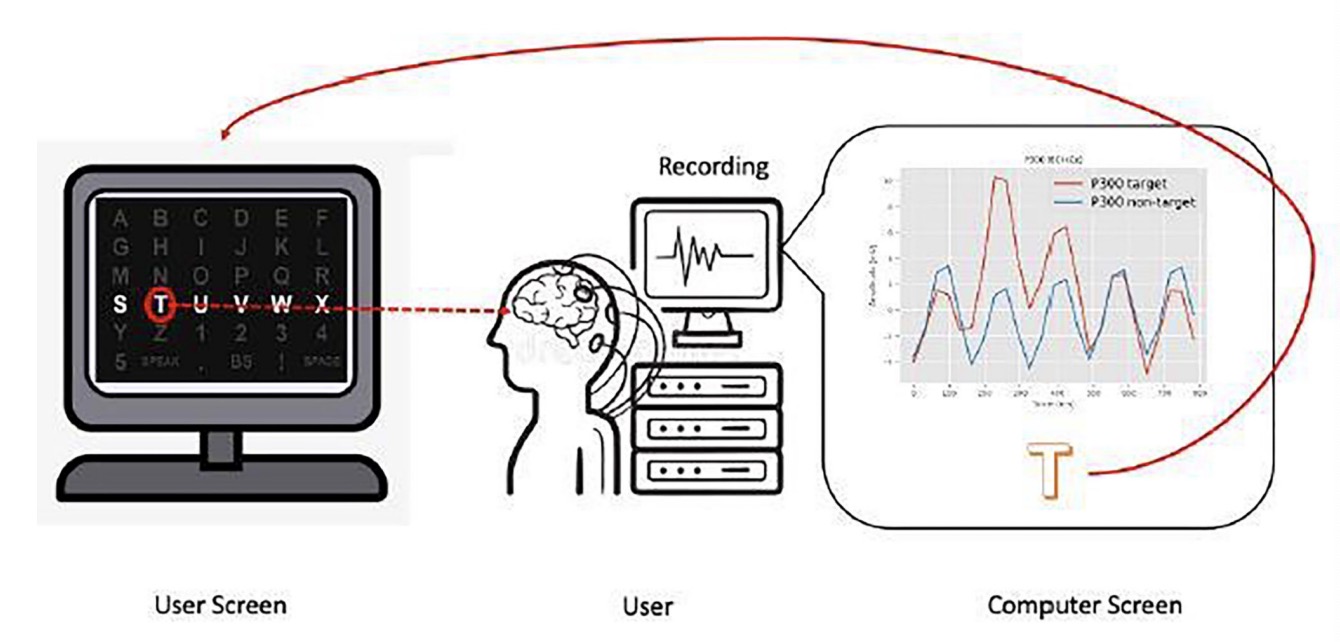}
\caption{An illustration of the P300 speller system \citep{ma2022bayesian}. At the beginning of the experiment, the user focuses the attention on the desired character or word they wish to type. The system presents a sequence of flashes on a virtual screen to the user, who responds to different flashes eliciting different brain signals (target or non-target). These EEG signals are captured by the electrodes and analyzed in a fixed time window after each flash to make a binary decision whether a P300 ERP response is produced. The binary classification results are then converted into character or word-level probabilities, and the one with the highest probability is shown on the screen.}
\label{fig:p300}
\end{figure}

The P300 speller is a type of BCI system that allows users to select characters or spell words on a computer screen without using a physical keyboard but instead the brain signals. It is based on the P300 event-related potential (ERP), which is a brain response, in the form of a specific pattern of voltage fluctuation, that occurs approximately 300 milliseconds after a relevant stimulus is presented. The stimuli are typically individual characters or symbols flashed on a computer screen in a grid-like layout, and ERP is detected and recorded by a scalp EEG cap or a similar device. After a stimulus is presented, the EEG signals captured by the electrodes are analyzed within a fixed time window by signal processing and machine learning algorithms, which detect the occurrence of the P300 response and determine the target character. Figure \ref{fig:p300} give a graphical illustration of the system. 

A key limitation of the existing system is that the stimuli are usually presented in a fixed and predetermined fashion. In addition, when presenting a word or a sentence, the system treats each character or each word independently, and totally ignores the inherent relations among the characters or words. As a result, a large number of stimulus flashes are usually required to achieve a certain level of accuracy of character or word identification. A language model essentially defines a collection of conditional probability distributions over the next token given the past tokens. Recently, pre-trained large language models (LLMs) such as GPT-3 \citep{brown2020language} have achieved striking success in natural language processing (NLP), and can produce coherent and human-like text. The objective of this article is to utilize language model as the prior information to improve the sampling efficiency of the P300 BCI system through an adaptive stimulus design. 

\paragraph{Contributions} Our contributions are three-fold:
\begin{itemize}
\item We introduce a novel sequential best-arm identification problem formulation that is motivated by a real-world BCI application. Each word the user wishes to type is treated as the optimal arm, and adaptive stimulus selection is the learning objective. By leveraging the pre-trained language model as an informative prior, the goal of the agent is to identify the target sequence of words as soon as possible (the fixed-confidence setting), or make as fewer mistakes as possible given a fixed number of flashes (the fixed-budget setting). 

\item We propose a sequential top-two Thompson sampling (\STTS) algorithm that utilizes the prior information in a coherent way. We derive the error probability bound in the fixed-budget setting that quantifies the prior effect through the conditional entropy of the prior distribution of the optimal arms. We also investigate the fixed-confidence setting theoretically. 

\item We conduct intensive experiments, using a P300 ERP-based BCI speller simulator \citep{ma2022bayesian}, along with the pre-trained GPT-2 \citep{radford2019language} and the OpenAI API for GPT-3 \citep{brown2020language}. We demonstrate the substantial improvement over several state-of-the-art baseline algorithms that do not use the prior information.
\end{itemize}

\subsection{Related work} 

We first review the literature on multi-armed bandits, then the literature on P300 BCI. 

For learning a single task, \citet{even2002pac} first introduced best-arm identification in the fixed-confidence setting, while \citet{audibert2010best} studied the fixed-budget setting. \citet{kaufmann2016complexity} investigated the optimal sample complexity, and \citet{jun2016top} explored the batch arm pulls setting. \cite{russo2016simple} proposed the top-two Thompson sampling as an effective anytime sampling rule that does not depend on the confidence parameter. Its theoretical properties were studied in \cite{russo2016simple, qin2017improving, shang2020fixed, qin2022adaptivity, jourdan2022top}. However, existing asymptotic analysis cannot demonstrate the prior effect. 

For learning multiple tasks sequentially, \citet{boutilier2020differentiable, simchowitz2021bayesian, kveton2021meta, azizi2022meta} studied meta-learning in the context of Bayesian bandits for cumulative and simple regret minimization. They assumed that an unknown instance prior is drawn from a known meta-prior. Then each task is sampled i.i.d.\ from this instance prior. In contrast, we assume the sequence of tasks is sampled from a joint prior distribution such that each task is \emph{not independent} of each other. This necessitates the prior-dependent analysis that has only been studied in the regret minimization setting \citep{russo2016information, hao2023leveraging}. 

For P300 BCI studies, there have recently emerged a number of proposals for adaptive stimulus selection.
\citet{speier2011natural} used language models to weigh the output of stepwise linear discriminant analysis (LDA) for classification confidence. \citet{park2012pomdp} framed the problem as a a partially observable Markov decision process (POMDP). However, the POMDP problem becomes difficult to solve for a real-time system with a large search space. \citet{ma2021adaptive} used Beta-Bernoulli Thompson sampling for adaptive stimulus selection, but did not formulate the problem as a best-arm identification, and only considered a single task. We refer to \cite{heskebeck2022multi} for a comprehensive review of the multi-armed bandits approaches in the BCI setting.

\section{Sequential best-arm identification}

We consider the problem that the agent sequentially interacts with $M$ bandit environments, with each interaction referred to as a task. In the P300 BCI setting, each task corresponds to a single word. Each environment, indexed by $m \in [M]$, is characterized by a random vector $\theta_m \in \mathbb R^J$ with a prior distribution $\nu_m(\cdot)$, which will be detailed in Section \ref{sec:prior_spec}. The action set is $\cA = \{a_1, \ldots, a_J\}\subseteq \mathbb R^J$, where $a_j$ is the standard basis vector. At each task $m$ and round $t$, the agent selects an action $A_{t,m}\in \cA$, and receives a reward $R_{t,m} = \langle A_{t,m}, \theta_m\rangle + \eta_{t,m}$, where $\langle \cdot, \cdot \rangle$ is the vector inner product, and $(\eta_{t,m})$ is a sequence of independent standard Gaussian random variables. The optimal arm, denoted as $A_m^* = \argmax_{a \in \cA} \theta_m^{\top} a$, is also a random variable. Moreover, let $(\Omega, \cF, \mathbb P)$ denote a measurable space.  Let $\cH_{t,m}$ denote the history of task $m$ up to round $t$, and $\cD_m=(\cH_{\tau_1, 1}, \ldots, \cH_{\tau_{m-1}, m-1})$. Write $\mathbb P_{t,m}(\cdot) = \mathbb P(\cdot|\cH_{t,m}, \cD_m)$. Let $[N] = \{1, 2, . . . , N\}$ for a positive integer $N$.

In the fixed-confidence setting \citep{even2002pac}, the agent chooses a policy $\pi^m = (\pi_{t,m})_{t=1}^{\infty}$ as the sampling rule. The horizon is not fixed in advance, however, as the agent decides a stopping time $\tau_m$ adapted to filtration, $\mathbb F^m = (\cF_{t,m})_{t=0}^{\infty}$, with $\cF_{t,m}=\sigma(A_{1,m}, R_{1,m}, \ldots, A_{t,m}, R_{t,m})$, where $\sigma$ is the Borel $\sigma$-algebra. At the end of the task, the agent takes an action or say a decision $\psi_m$. In the P300 BCI example, this decision is the word the system believes the user intends to type. For a given a confidence level $\delta \in (0, 1)$, the objective is to output a sequence of arms that are optimal for each task with probability at least $1-\delta$ as soon as possible. 

In the fixed-budget setting \citep{bubeck2009pure}, the agent is given a budget $n$ for each task, choose a policy $\pi^m = (\pi_{t,m})_{t=1}^{n}$, and takes an action  $\psi_m$ at the end of the task. The objective is to make the cumulative probability that $\psi_m$ is sub-optimal as small as possible.

\subsection{Prior specification}
\label{sec:prior_spec}

A language model $\rho$ defines a collection of conditional probability distributions $(\rho_1,\ldots, \rho_{M})$, where $\rho_m$ denotes a probability distribution over the $m$th word given the first $m-1$ words. In a P300 BCI experiment, the word that an individual attempts to type is viewed as the optimal arm. The joint distribution over the collection of optimal arms $(A_1^*,\ldots, A_M^*)$ can be written through the chain rule:
\begin{equation}\label{eqn:prior_optimal_arm}
 \rho(A_1^*,\ldots, A_M^*)=\prod_{m=1}^M\rho_m = \prod_{m=1}^M\mathbb P\left(A_m^*=\cdot |A_1^*, \ldots, A_{m-1}^*\right)\,.
\end{equation}
For each task, the prior distribution of $\theta_m^*$ can be specified through
\begin{equation}\label{eqn:prior_specification}
\begin{split}
\nu_m&=  \mathbb P\left(\theta_m\in\cdot|A_1^*,\ldots, A_{m-1}^*\right)\\
&=\sum_{j=1}^J\underbrace{\mathbb P\left(\theta_m\in\cdot|A_m^*=j, A_1^*,\ldots, A_{m-1}^*\right)}_{\text{prior of the conditional mean reward}}\underbrace{\mathbb P\left(A_m^*=j|A_1^*,\ldots, A_{m-1}^*\right)}_{\text{prior of the optimal arm}}\,,
\end{split}
\end{equation}
which is a mixture distribution. While the prior of the optimal arm can be defined by \eqref{eqn:prior_optimal_arm}, there are several ways to specify the prior of the conditional mean reward, depending on the problem setting. For the P300 BCI example, the reward only differs upon whether the stimulus is a target or a non-target. Consequently, all sub-optimal arms share the same mean reward. Thus it is natural to assume that, conditional on $A_m^*=j, A_1^*,\ldots, A_{m-1}^*$, is of the form, 
\begin{equation} \label{eqn:prior_mean_reward}
\theta_m | A_m^*=j, A_1^*,\ldots, A_{m-1}^* \,{\buildrel d \over =}\, (\mu, \ldots, \underbrace{\mu+\Delta}_{j\text{th}}, \ldots, \mu),
\end{equation}
where $\mu\sim \cN(0, \sigma_0^2)$, and $\Delta\sim \exp(\sigma_1)$, for some $\sigma_0,\sigma_1 > 0$.

\section{Sequential top-two Thompson sampling}

We propose a sequential top-two Thompson sampling (\STTS) algorithm that utilizes the prior information in a coherent way. We also develop the corresponding stopping rule and the decision rule for both the fixed-confidence setting and the fixed-budget setting.

\subsection{Sampling procedure}

\STTS \ assumes there exists a posterior sampling oracle that can be obtained exactly when a conjugate prior is used, or through various approximation methods, such as Markov chain Monte Carlo.  

\begin{definition}[Posterior sampling oracle]
Given a prior $\nu_m$ over $\theta_m$ and history $\cH_{t,m}, \cD_m$, the posterior sampling oracle, \textit{SAMP}, is a subroutine which returns a sample from the posterior distribution $\mathbb P_{t,m}(\theta_m\in\cdot)$. Multiple calls to the procedure result in independent samples.
\end{definition}

Our proposed \STTS \ algorithm is an extension of the top-two Thompson sampling \citep{qin2022adaptivity, russo2016simple} that sequentially calls the language model $\rho$ to construct an informative prior. At task $m$ and round $t$, \STTS \ first draws a posterior sample $\tilde \theta_{t,m}$ using \textit{SAMP} as well as the language model $\rho$, and set $A_{t,m, 1} = \argmax_{a \in \cA} \tilde \theta_{t,m}^{\top} a$. Then \STTS \  repeatedly samples from \textit{SAMP} to obtain  $\tilde \theta_{t,m}$, and set $A_{t,m,2} = \argmax_{a \in \cA} \tilde \theta_{t,m}^{\top} a$ until $A_{t,m,1} \neq A_{t,m,2}$. We pick $A_{t,m} = A_{t,m,1}$ with probability equal to $\beta\in (0, 1]$, and $A_{t,m} = A_{t,m, 2}$ with probability equal to $1-\beta$. In practice, we recommend $\beta=1/2$, following  \cite{qin2022adaptivity}.

\subsection{Stopping rule and decision rule}
\label{sec:stopping_rule}

For the stopping rule in the fixed-confidence setting, we employ the Chernoff stopping rule introduced by \cite{garivier2016optimal, shang2020fixed}. Let $N_{t, a_i}$ denote the number of pulls of arm $a_i$ before round $t$ for each task, and $\mu_{t, a_i}$ the posterior mean for arm $a_i$ and $\mu_{t, a_i, a_j} = (N_{t, a_i}\mu_{t, a_i}+N_{t, a_j}\mu_{t, a_j})/(N_{t, a_i}+N_{t, a_j})$. We further define $Z_{t, a_i, a_j}=N_{t, a_i}\text{KL}(\mu_{t, a_i}, \mu_{t, a_i, a_j})$ for any two arms $a_i$ and $a_j$, such that $Z_{t}(a_i, a_j)=0$ if $\mu_{t, a_j}\geq \mu_{t, a_i}$, and $Z_{t}(a_i, a_j)=Z_{t, a_i, a_j}+ Z_{t, a_j, a_i}$ otherwise, where $\text{KL}(\mu_1, \mu_2)$ is the KL-divergence between two distributions with mean $\mu_1$ and $\mu_2$. For each task $m$, the Chernoff stopping rule is, 
\begin{equation}\label{eqn:stopping_rule}
\tau_m = \inf\left\{t\in\mathbb N: \max_{a_i\in\cA}\min_{a_j\in\cA\setminus \{a_i\}} Z_{t}(a_i, a_j) \ge \gamma_{t,\delta}\right\}\,,
\end{equation}
where $\gamma_{t,\delta}$ is the threshold parameter and $\delta$ controls the level of confidence. As noted in \cite{shang2020fixed}, $Z_t(a_i, a_j)$ can be interpreted as a generalized likelihood ratio statistic. As the agent interacts with $M$ tasks sequentially, we employ the Bonferroni correction to handle multiple comparisons \citep{dunn1961multiple}. As such, the family-wise error rate $\delta_M = \delta/M$,  and the level of confidence for each task becomes $1 - \delta/M$. For the final decision rule, we choose the Bayes optimal decision rule, $\psi_{m} = \argmax_{i}\mu_{\tau_m, a_i}$.

We summarize the full procedure for the fixed-confidence setting in Algorithm \ref{alg:cap}. For the fixed-budget setting, the only change is that we stop the algorithm when some pre-specified budget constraint is met. 

\begin{algorithm}
\caption{\STTS \ for sequential best-arm identification}\label{alg:cap}
\begin{algorithmic}
\INPUT LLM $\rho$, sampling oracle \textit{SAMP}, level of confidence $\delta$, sampling parameter $\beta$
\OUTPUT A sequence of recommended actions $\{\psi_1,\ldots,\psi_{M}\}$
\FOR{$m \in [M]$}
\STATE Construct the informative prior based on $\rho$ and history $\{\psi_1,\ldots,\psi_{m-1}\}$
\STATE Set $t=0$
\WHILE{TRUE}
\STATE Set $t = t+1$
\STATE Sample $\tilde \theta_{t,m}$ from \textit{SAMP} and set $A_{t,m, 1} =A_{t,m, 2} =\argmax_{a}\tilde \theta_{t,m}^{\top}a$
\WHILE{$A_{t,m, 1} =A_{t,m, 2} $}
\STATE Re-sample $\tilde \theta_{t,m}$ from \textit{SAMP} and set $A_{t,m,2} =\argmax_{a}\tilde \theta_{t,m}^{\top}a$
\ENDWHILE
\STATE Sample $C_{t,m} \sim \text{Bernoulli}(\beta)$
\STATE Pull arm $A_{t,m} = A_{t,m, 1} C_{t,m} +A_{t,m, 2}  (1-C_{t,m})$ and receive $R_{t,m}$
\IF{$\max_{a_i\in\cA}\min_{a_j\in\cA\setminus \{a_i\}} Z_{t}(a_i, a_j) \ge \gamma_{t,\delta_M}$}
\STATE Identify the best arm $\psi_m$ 
\STATE STOP and break the while loop
\ENDIF
\STATE Update \textit{SAMP} 
\ENDWHILE
\ENDFOR
\end{algorithmic}
\end{algorithm}

\section{Theoretical results}

We first consider the fixed-budget setting. When the recommended word at the end of one task is wrong, we say the agent makes a mistake. The next theorem derives the corresponding error probability bound.

\begin{theorem}\label{thm:fixed_budget}
Consider a sequence of $M$ best-arm identification problems, and assume for each task $m$, all the sub-optimal arms have the same and known sub-optimality gap,  $\Delta_m=\langle A_{m}^*-a, \theta_m\rangle$, for any $a\in\cA$. If \STTS \ is applied with $\beta\geq 1/2$ and with the Bayes optimal decision rule $\psi_{m}$, then for any positive integer-valued budget $n$,
\begin{equation*}
\begin{split}
\frac{1}{M}\sum_{m=1}^M\mathbb P\left(\psi_m\neq A_m^*\right)& \underbrace{\leq \sum_{m=1}^M\frac{\sqrt{\mathbb H(A_m^*|A_{m-1}^*,\ldots, A_1^*)}}{\Delta_m}\frac{6}{M}\sqrt{\frac{\log(J(1+n))}{1+n}}}_{\text{main term}}\\
&+\underbrace{1-\frac{1}{M}\sum_{m=1}^M\prod_{j=1}^{m-1}\left(1-\frac{6}{\Delta_m}\sqrt{\frac{\log(J(1+n))\mathbb H(A_m^*|A_{m-1}^*,\ldots, A_1^*)}{1+n}}\right)}_{\text{remainder term}}\,,
\end{split}
\end{equation*}
where $\mathbb H(\cdot|\cdot)$ is the conditional entropy. 
\end{theorem}

\begin{figure}
\centering
\begin{tabular}{cc}
\includegraphics[width=5cm,height=3.5cm]{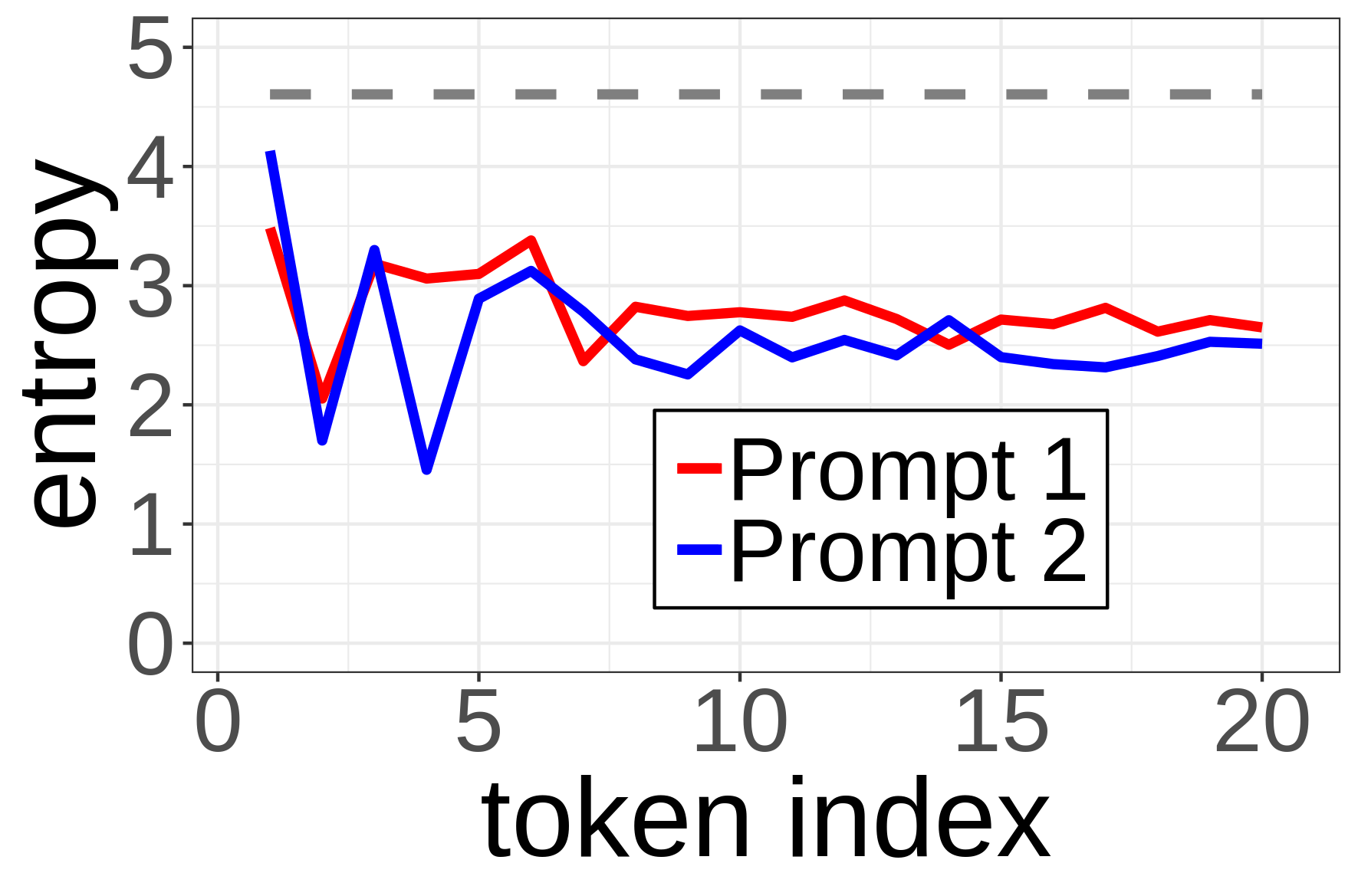} &
\includegraphics[width=5cm,height=3.5cm]{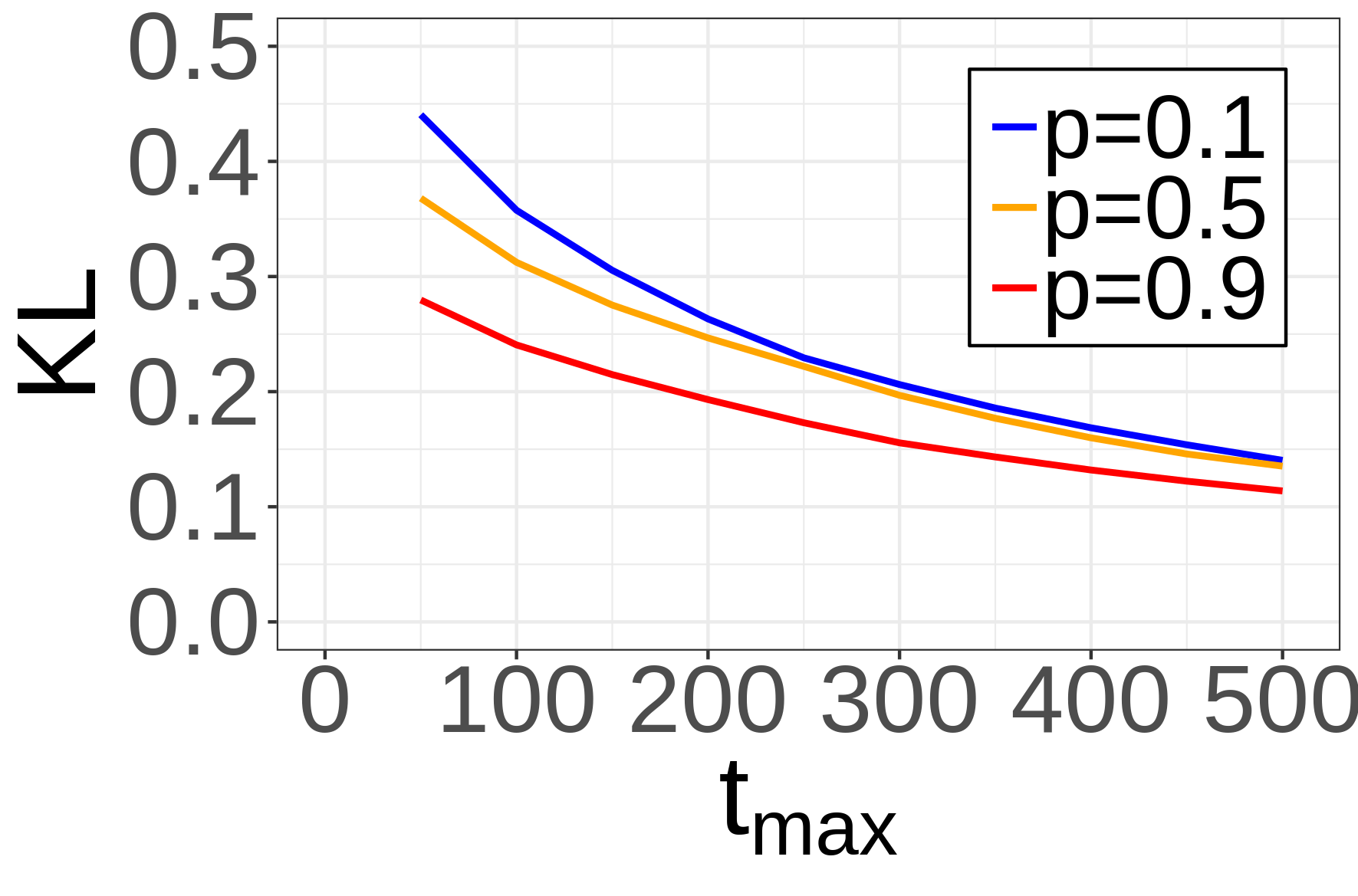}
\end{tabular}
\caption{Left panel: the entropy of the next word under two different prompts (red and blue curves) versus and the entropy of a uniform distribution (the dashed line). Right panel: the KL divergence between the allocation rule of \STTS \ and the optimal allocation rule. A larger $p$ indicates a stronger prior. When \STTS \ reaches $t_{\max}$, it stops.}
\label{fig:allocation}
\end{figure}

In this bound, the term $\mathbb H(A_m^*|A_{m-1}^*,\ldots, A_1^*)$ characterizes the effect of the prior. In P300 BCI, the prior distribution of $A_m^*$ is informed by a language model such that $\mathbb H(A_m^*|A_{m-1}^*,\ldots, A_1^*)$ is much smaller than $\log(J)$. This is validated by our experiment in Figure \ref{fig:allocation}, left panel, where the entropy of the probability distribution over next word outputted by GPT-2 \citep{radford2019language} is shown to be much smaller than an uniform distribution. The remainder term is the price the agent pays for making mistakes. When the budget $n$ for each task tends to infinity, the remainder term goes to 0. 

There is another practical scenario where at the end of task $m$, the agent is given the identity of the optimal arm $A_m^*$. If the recommended action is wrong, the agent pays an extra price $c$. In P300 BCI, if the system outputs a wrong recommendation, the participant will gaze at the backspace and the system will repeat the process until the right word is recommended. In this case, the error probability can be bounded by
\begin{equation*}
\frac{1}{M}\sum_{m=1}^M \mathbb P\left(\psi_m\neq A_m^*\right) \leq\sum_{m=1}^M\frac{\sqrt{\mathbb H(A_m^*|A_{m-1}^*,\ldots, A_1^*)}}{\Delta_m}\frac{6}{M}\sqrt{\frac{\log(J(1+n))}{1+n}},
\end{equation*}
and the number of mistakes is bounded by $c\sum_{m=1}^M \mathbb P\left(\psi_m\neq A_m^*\right)$.

In this theorem, we have assumed the same and known gaps for all sub-optimal arms, which allows us to use simple regret guarantee to bound the error probability. For P300 BCI, this is a reasonable assumption, since all the non-target stimuli have the same levels of EEG responses and the gap can usually be estimated though the offline data \citep{ma2021adaptive}. 

We next consider the fixed-confidence setting. Choosing the threshold $\gamma_{t, \delta}=4\ln(4+\ln(t))+2C(\ln((J-1)/\delta)/2)$ for some constant $C$, and applying Theorem 1 in \cite{shang2020fixed} leads to an asymptotic sample complexity bound for \STTS, coupled with the Bonferroni correction, as, 
\begin{equation*}
\limsup_{\delta\to 0}\frac{\sum_{m=1}^M\mathbb E[\tau_m]}{\log(1/\delta)} \leq \sum_{m=1}^M\int\frac{2J}{\Delta_m}  \d\nu_m(\Delta_m)\,,
\end{equation*}
where we take the expectation over the prior of $\theta_m$ on both sides. However, this asymptotic result can not fully characterize the prior effect since the asymptotic complexity measure only depends on the prior of the gap $\Delta_m$, rather than the prior of the position of the optimal arm. Typically, the overall sample complexity consists of two parts: the cost due to the optimal allocation rule, and the cost due to finding the optimal allocation rule. The latter is a lower-order term with respect to $\log(1/\delta)$. We conjecture that a more informative prior could greatly reduce the cost of finding the optimal allocation rule in the finite time. We conduct a numerical experiment to verify our conjecture. The detailed simulation setting is given in Appendix \ref{sec:allocation}. We report in Figure \ref{fig:allocation}, right panel, the KL-divergence between the asymptotically optimal allocation rule and the allocation rule induced by \STTS. It is clearly seen that, when the prior is stronger, as reflected by a larger value of $p$ as defined in Appendix \ref{sec:allocation}, the allocation rule induced by \STTS \ converges faster to the asymptotically optimal allocation rule. Meanwhile, a rigorous analysis requires a much more involved finite-time problem dependent analysis for top-two Thompson sampling, and we leave it as future research.

\section{Synthetic experiments}
\label{sec:synthetic}

We first carry out synthetic data experiments to investigate the empirical performance of our method and study the effect of the prior. We compare with the following baselines solutions. 

\begin{itemize}
\item Vanilla top-two Thompson sampling (\VTTS) \citep{russo2016simple, qin2022adaptivity}: top-two Thompson sampling that does not use any informative prior from the LLM.
\item Batch Racing (\BR) \citep{jun2016top}: A frequentist algorithm that uses confidence interval to identify best arms in the fixed-confidence setting.
\item Random Policy (\Random): A uniform sampling rule.
\end{itemize}

\VTTS \ and \Random \ use the same stopping rule and decision rule as described in Section \ref{sec:stopping_rule}, while \BR \ is only for the fixed-confidence setting. There are several other popular best-arm identification algorithms \citep{jamieson2014best, garivier2016optimal} in the literature, but none of them is designed for the sequential task setting. 

We define the prior for the mean reward $\theta_m^*$ through \eqref{eqn:prior_specification} that requires the specification for the prior of the conditional mean reward and the prior of the optimal arm. We assume the prior of the optimal arms satisfies the Markov property, such that the distribution of the $m$th optimal arm depends only on the $(m-1)$th optimal arm, instead of the entire history of optimal arms. Specifically, the optimal arm for $m$th  task is sampled from
\begin{equation}\label{eqn:prior_LLM}
\PP(A^*_{m} = j)=  
\begin{cases}
p & \text{if} \ j \in \{A^*_{m-1}+1, A^*_{m-1}+1-J\} \cap [J]\\
(1-p)/(J-1) &\text{if} \ j \in \{A^*_{m-1}+1, A^*_{m-1}+1-J\}^c \cap [J]\,,
\end{cases}
\end{equation}
where $p$ is a parameter that controls the strength of the prior, and $S^c$ denotes the complement set of $S$. In practice, there are often offline data available \citep{ma2021adaptive}, such that $\Delta$ can be well-estimated. Thus, the prior of the conditional mean reward follows $\theta_m|A_m^*=j, A_1^*,\ldots, A_{m-1}^* \sim \cN((\mu, \ldots, \mu+\Delta, \ldots, \mu), \sigma_0^2 \bI_J)$ for any $j\in[J]$. 

We also note that \STTS \ is a general algorithm that can be coupled with any prior specification. As an illustration, we also consider a Gaussian prior specification in Appendix \ref{sec:alter_prior}.

\subsection{Fixed-confidence setting}

We begin with deriving the posterior distribution used by different variants of top-two Thompson sampling. In our setting, the prior of the mean reward of the $j$th arm follows a Gaussian mixture distribution, 
\begin{equation}\label{eqn:prior_LLM}
\mathbb P(\theta_{m,j} \in \cdot|A_1^*,\ldots,A_{m-1}^*) =  p_{m,j} \cN(\mu+\Delta, \sigma^2_{0}) + (1-p_{m,j}) \cN(\mu, \sigma^2_{0}),
\end{equation} 
and $p_{m,j} \in [0,1]$ is specified later. Due to the Gaussian reward noise, the posterior distribution remains a Gaussian mixture distribution that can be efficiently sampled from. As commented earlier, when there is an oracle or external resource that reveals the identity of the optimal arm at the end of each task, \STTS \ can start with an exact prior, which we call \texttt{STTS-Oracle}. We set $p_{m,j}$ for different variants of the algorithm as follows:
\begin{itemize}
\item For \texttt{STTS-Oracle}, we set $p_{m,j} = \PP(A^*_m = j|A^*_{m-1}=a_{m-1}^*)$ for $j \in [J]$, where $\{a_1^*, \ldots, a_{m-1}^*\}$ denotes the realized instances of the optimal arms.
\item For \STTS, we set $p_{m,j} = \PP(A^*_m = j|A^*_{m-1} = \psi_{m-1})$ for $j \in [J]$, where $\{\psi_1, \ldots, \psi_{m-1}\}$ are the recommended actions. 
\item For \VTTS, we set $p_{m,1}=\ldots=p_{m,J} = J^{-1}$.
\end{itemize}
Let $\mu_{m,t}$ denote the posterior mean, and $\sigma_{m,t,j}^2$ the posterior variance. Then the Chernoff stopping rule defined in \eqref{eqn:stopping_rule} becomes
\begin{equation}\label{eqn:stopping_rule_chernoff}
\tau_m = \min\left\{t:\min_{j\neq \psi_{t,m}}\frac{\mu_{m, t,j_1}-\mu_{m, t,j_2}}{\sqrt{\sigma_{m, t,j_1}^2+\sigma_{m, t,j_2}^2}}\geq \gamma_t\right\},
\end{equation}
where $\psi_{t,m} = \argmax_a a^{\top}\mu_{m, t}$, and$\gamma_t = \sqrt{2 \log(\log(t)M/\delta)}$. Here, we approximate the KL-divergence of two Gaussian mixture distributions by the KL-divergence of two Gaussian distributions with the same mean and variance \citep{hershey2007approximating}. As the theoretical stopping rule of \BR \ is rather conservative, we multiply its range by a factor of 0.25.

\begin{figure}[t!]
\centering
\begin{tabular}{cc}
\includegraphics[width=5cm,height=3.5cm]{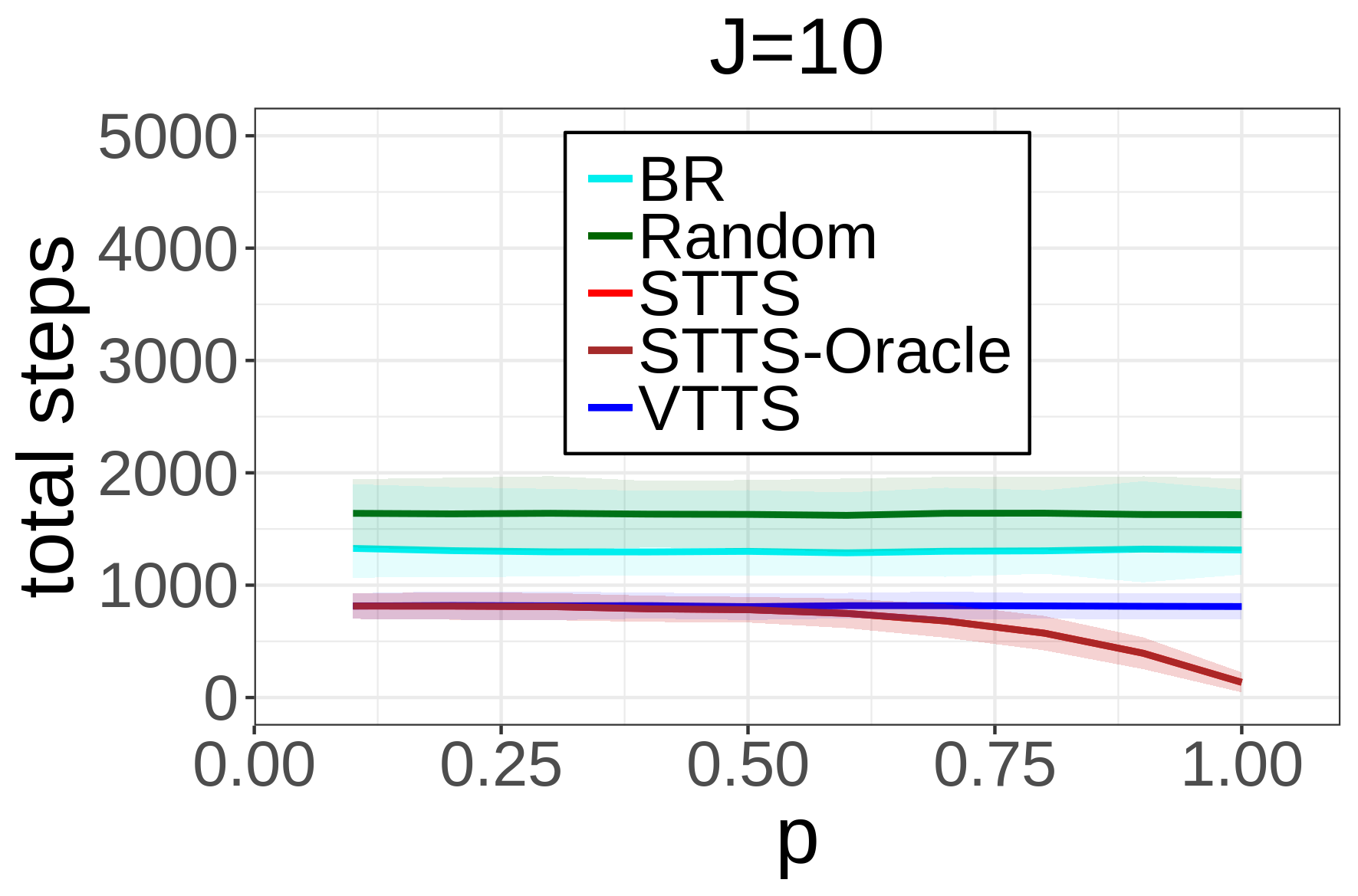}
\includegraphics[width=5cm,height=3.5cm]{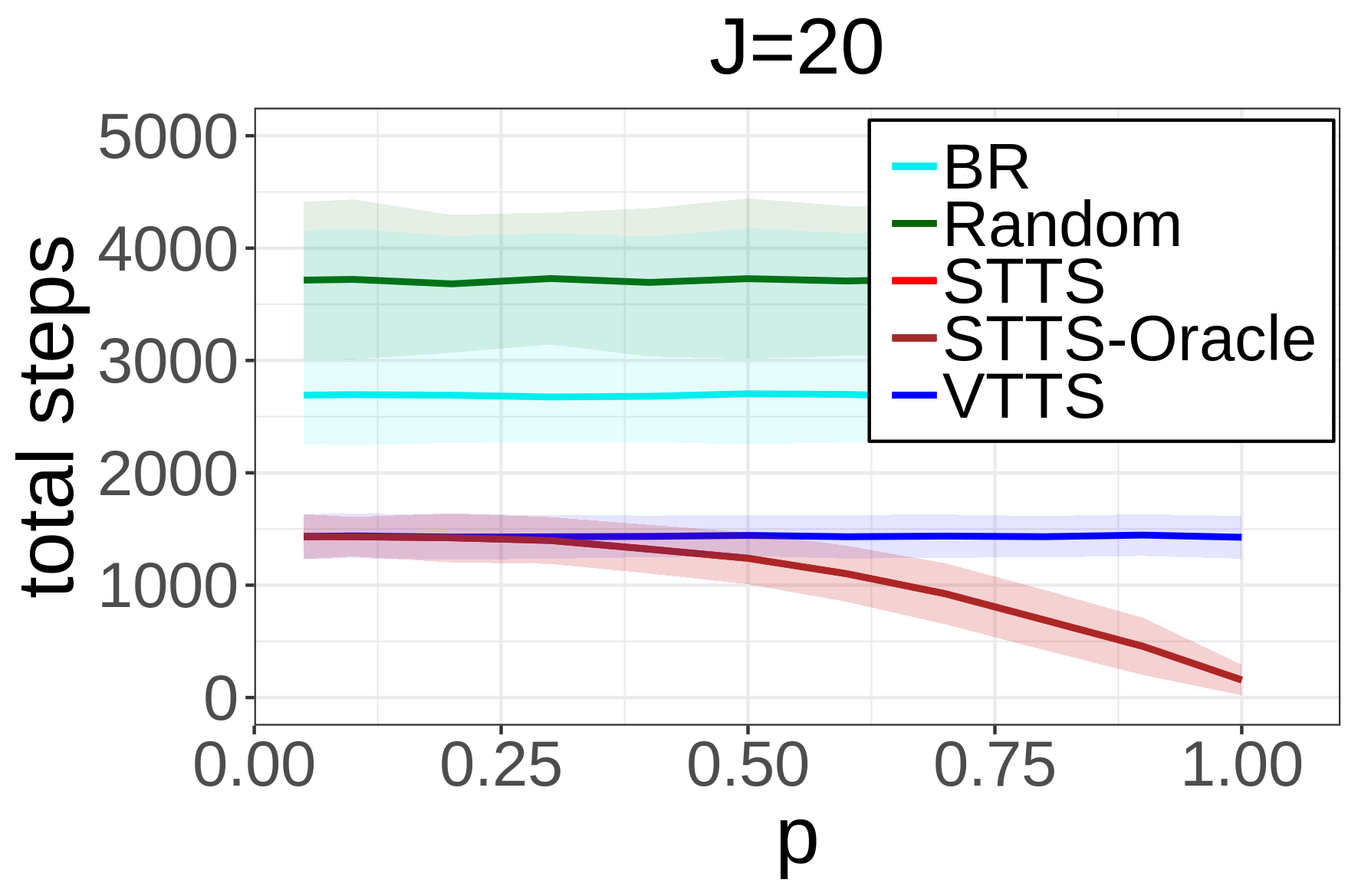}
\end{tabular}
\caption{The empirical performance for the fixed-confidence setting. 
}
\label{fig:FC_synthetic} 
\end{figure}

We take one sentence as one experiment, and each sentence consists of $M=20$ words or tasks. We set the confidence level $\delta=0.1$, and the prior parameters $\mu=0, \sigma_0^2=0.2, \Delta=2$ in \eqref{eqn:prior_LLM}. We vary the number of arms $J \in \{10,20\}$, and $p \in \{J^{-1}\} \cup \{0.1,0.2,\ldots,1.0\}$. A larger $p$ indicates a stronger prior effect, and $p=J^{-1}$ means a non-informative prior. We replicate each experiment $B=200$ times. 

We consider two accuracy measures for each experiment, the 0-1 accuracy, where as long as the agent makes at least one mistake among 20 words or tasks, we mark that experiment a failure, and the average accuracy, where we compute the percentage of correctly identified words among a sentence: $\sum_{m=1}^M\sum_{b=1}^B\mathbb I\{\psi_m^b=A_{m,b}^*\}/(BM)$, where $\psi_m^b, A_{m,b}^*$ denotes the recommended action and the optimal action for the $m$th task of the $b$th experiment, respectively. .  

Figure \ref{fig:FC_synthetic} reports the total sample complexity under a varying $p$, while the accuracy of all methods exceed $1-\delta=0.9$. It is seen from the plot that, with the prior becoming more informative, \STTS \ clearly outperforms \VTTS, \BR \, and \Random. The improvement also increases with an increasing number of arms. When the prior of the optimal arms is uniformly distributed, \STTS \ coincides with \VTTS \ as expected. Furthermore, \STTS \ and \texttt{STTS-Oracle} exhibit similar performances. This is because, as the accuracy of \STTS \ is consistently close to one, $\PP(A^*_m = \cdot|A^*_{m-1} = a_{m-1}^*)$ and $\PP(A^*_m = \cdot|A^*_{m-1} = \psi_{m-1})$ are almost identical.

\subsection{Fixed-budget setting} 
\label{subsec:syn_fix_bud}

\begin{figure}[b!]
\centering
\includegraphics[width=3.4cm]{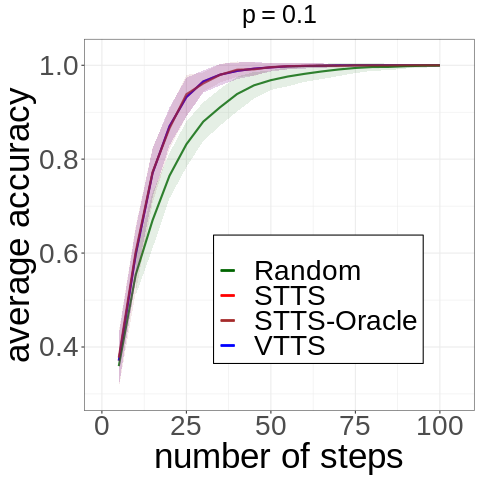}
\includegraphics[width=3.4cm]{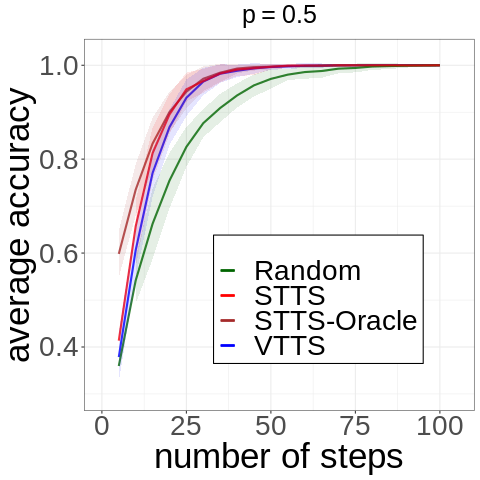}
\includegraphics[width=3.4cm]{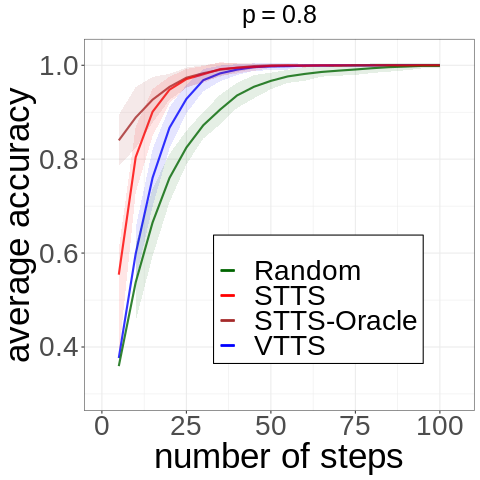}
\includegraphics[width=3.4cm]{SynP300_acc_J=10_p=J_1.png}
\caption{The empirical performance for the fixed-confidence setting with $J=10$. 
}
\label{fig:FB_synthetic1}
\end{figure}

We next consider fixed-budget setting. The sampling and decision rules remain the same while the stopping rule is determined by whether it reaches the pre-specified budget of maximum step $t_{\max}$ per task, which varies among $\{5, 10, 15, \ldots, 100\}$. We set $J \in \{10, 20\}$, $p \in \{0.1, 0.5, 0.8, 1.0\}$, and replicate each experiment $B=200$ times. Figure \ref{fig:FB_synthetic1} reports the average accuracy over all tasks for $J=10$. We report the results for $J=20$ in Appendix \ref{subsec:add_exp1}. It is seen from the plot that, when the prior becomes more informative, \STTS \ and \texttt{STTS-Oracle} reach a high accuracy with a smaller budget. When the prior is non-informative, \STTS\ and \texttt{STTS-Oracle} perform similarly.

\section{P300 BCI experiments}

We carry out a P300 BCI experiment using the speller simulator \citep{ma2022bayesian}, in conjunction with an open-source GPT-2 \citep{radford2019language} and the OpenAI API for GPT-3  \citep{brown2020language}.

\subsection{Experiment setup}
\label{subsec:realexp_setup}

We simulate the brain EEG signals using the P300 ERP-based BCI speller simulator of \citet{ma2022bayesian}. Specifically, we set the number of electrodes to 16, the noise variance $\sigma^2_{\text{EEG}} \in \{1,2.5\}$, the noise spatial correlation based on a Gaussian kernel function, the noise temporal correlation from an $AR(1)$ model with an autocorrelation $0.9$, and the mean magnitude of the target stimulus five times that of the non-target stimulus. In accordance with the current practice \citep{manyakov2011comparison}, we first train a binary classifier for the P300 offline data based on stepwise linear discriminant analysis \citep{donchin2000mental, krusienski2008toward}. This classifier converts the raw EEG signals into the classifier scores, which are taken as the rewards in our setting. A higher score indicates that the EEG signal is more likely to correspond to a target stimulus.

Pre-trained large language models (LLMs) such as GPT-2 or GPT-3 can produce coherent and human-like text. In our experiment, we use GPT-3 to generate the sentence of words that a participant wishes to type, and use GPT-2 to inform the prior probability distribution. This also mimics the potentially imperfect prior information that commonly appears in the real-world scenarios. Specifically, we use GPT-3 to generate $M=20$ words given a prompt. These are the words that the participant wishes to type and form the optimal arms $\{A^*_1,\ldots,A^*_{M}\}$. We consider two prompts: Prompt 1 \emph{(``The most popular food in the United States is'')}, and Prompt 2 \emph{(``My favorite sports is'')}. We repeat each prompt $B=100$ times. Following the top-$K$ sampling \citep{fan2018hierarchical} and the nucleus sampling \citep{holtzmancurious}, we truncate the vocabulary size of GPT-2 from the original size of 50257 to 100, so keeping the candidate words with the top-100 highest probabilities. This effectively reduces the size of action space $J$ to 100. We specify the prior of the optimal arm using the probability distribution informed by GPT-2. We set the confidence level at $1-\delta=0.9$.

\subsection{Experiment results} 
\label{subsec:realdata_res}

We consider two cases. First, if an algorithm recommends a wrong action, we stop the experiment, record this experiment as a failure, and compute the 0-1 accuracy. Second, if an algorithm recommends a wrong action, we let the experiment continue to run, but reveal the identity of the optimal arm, and compute the average accuracy.  We also compare with Beta-Bernoulli Thompson sampling (\BBTS) of \cite{ma2021adaptive} for adaptive stimulus selection.  Since there is no theoretical justification of the $p_{\max}$ in the stopping rule of \BBTS, we choose it in a heuristic way as $p_{\max} = 1 - \delta/(1000M)$. 

\begin{table}[t!]
\centering
\begin{tabular}{|c|cc|cc|} 
\hline
Method & $\sigma^2_{\text{EEG}}$  & total steps (std)& $\sigma^2_{\text{EEG}}$  & total steps (std) \\
\hline
\STTS&1&668.30 (136.5)&2.5&1623.0 (240.6)\\
\VTTS&1&1445.0 (237.2)&2.5&2493.2 (275.5)\\
\Random&1&6865.5 (826.6)&2.5&13523. (829.9)\\
\BBTS&1&1430.4 (203.5)&2.5&1712.5 (248.1)\\
\BR&1&2611.8 (112.6)&2.5&4396.4 (322.8)\\
\hline
\end{tabular}
\caption{Total number of steps for Prompt 1.}
\label{tab:FC_food}
\vspace{-0.3in}
\end{table}

\textbf{Fixed-confidence setting.}
We stop the algorithm when the Chernoff stopping rule \eqref{eqn:stopping_rule_chernoff} is satisfied. Table \ref{tab:FC_food} reports the total number of steps, i.e., flashes, while the 0-1 accuracy and the average accuracy for all algorithms are always above 0.9. It is seen from the table that, facilitated by the LLM-informed prior, \STTS \ can reduce the total number of stimulus flashes by $25\%$ to $50\%$ while maintaining about the same accuracy. 
\begin{figure}[b]
\centering
\includegraphics[width=3.4cm]{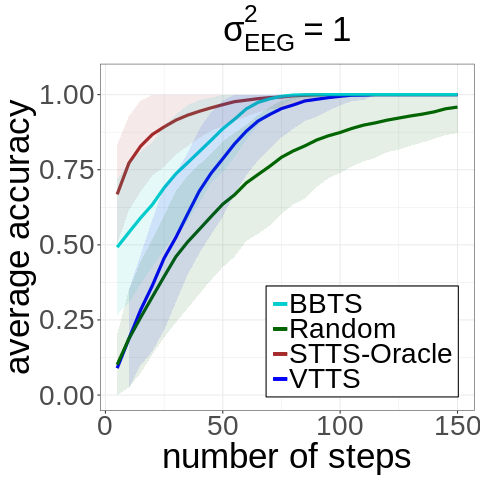}
\includegraphics[width=3.4cm]{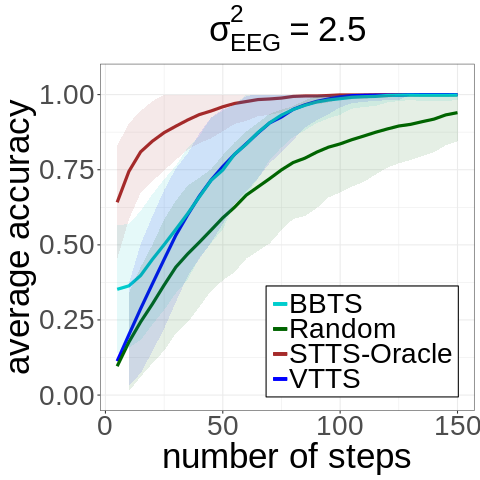}
\includegraphics[width=3.4cm]{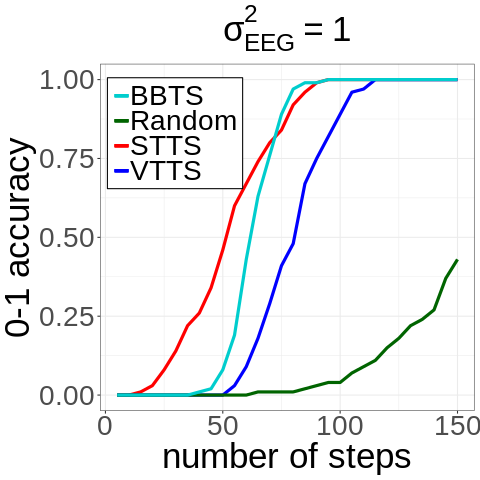}
\includegraphics[width=3.4cm]{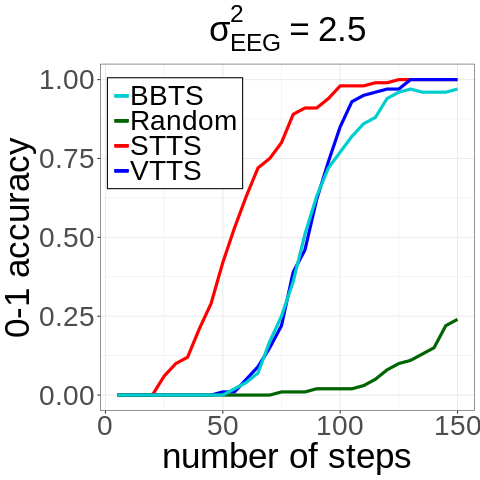}
\caption{The average accuracy and 0-1 accuracy in the fixed budget setting.}
\label{fig:FB_real_case1} 
\end{figure}

\textbf{Fixed-budget setting.}
We stop the algorithm stops when it reaches a pre-specified maximum number of flashes $t_{\max} \in \{5,10,15,\ldots,150\}$. Figure \ref{fig:FB_real_case1} reports both the 0-1 accuracy and the average accuracy. It is seen from the plot that\texttt{STTS-Oracle} \ achieves the highest average accuracy, while \STTS \ achieves the highest 0-1 accuracy with a smaller budget than the competing methods.

\section{Conclusion}

The BCI technology has the potential to revolutionize human communication by bypassing physical constraints and interacting directly with the brain. However, a critical challenge lies in its low sampling efficiency. We propose a sequential best-arm identification formulation for the P300 BCI system that greatly enhances its sampling  efficiency. We hope our work could open the door of using adaptive bandit algorithms for this important application. There should be no ethics/societal risk for this work since all numerical analyses in this article are based on simulations. 

\bibliographystyle{plainnat}
{\small
\bibliography{ref}

\begin{thebibliography}{33}
\providecommand{\natexlab}[1]{#1}
\providecommand{\url}[1]{\texttt{#1}}
\expandafter\ifx\csname urlstyle\endcsname\relax
  \providecommand{\doi}[1]{doi: #1}\else
  \providecommand{\doi}{doi: \begingroup \urlstyle{rm}\Url}\fi

\bibitem[Audibert et~al.(2010)Audibert, Bubeck, and Munos]{audibert2010best}
Jean-Yves Audibert, S{\'e}bastien Bubeck, and R{\'e}mi Munos.
\newblock Best arm identification in multi-armed bandits.
\newblock In \emph{COLT}, pages 41--53, 2010.

\bibitem[Azizi et~al.(2022)Azizi, Kveton, Ghavamzadeh, and
  Katariya]{azizi2022meta}
Mohammadjavad Azizi, Branislav Kveton, Mohammad Ghavamzadeh, and Sumeet
  Katariya.
\newblock Meta-learning for simple regret minimization.
\newblock \emph{arXiv preprint arXiv:2202.12888}, 2022.

\bibitem[Boutilier et~al.(2020)Boutilier, Hsu, Kveton, Mladenov, Szepesvari,
  and Zaheer]{boutilier2020differentiable}
Craig Boutilier, Chih-Wei Hsu, Branislav Kveton, Martin Mladenov, Csaba
  Szepesvari, and Manzil Zaheer.
\newblock Differentiable meta-learning of bandit policies.
\newblock \emph{Advances in Neural Information Processing Systems},
  33:\penalty0 2122--2134, 2020.

\bibitem[Brown et~al.(2020)Brown, Mann, Ryder, Subbiah, Kaplan, Dhariwal,
  Neelakantan, Shyam, Sastry, Askell, et~al.]{brown2020language}
Tom Brown, Benjamin Mann, Nick Ryder, Melanie Subbiah, Jared~D Kaplan, Prafulla
  Dhariwal, Arvind Neelakantan, Pranav Shyam, Girish Sastry, Amanda Askell,
  et~al.
\newblock Language models are few-shot learners.
\newblock \emph{Advances in neural information processing systems},
  33:\penalty0 1877--1901, 2020.

\bibitem[Bubeck et~al.(2009)Bubeck, Munos, and Stoltz]{bubeck2009pure}
S{\'e}bastien Bubeck, R{\'e}mi Munos, and Gilles Stoltz.
\newblock Pure exploration in multi-armed bandits problems.
\newblock In \emph{Algorithmic Learning Theory: 20th International Conference,
  ALT 2009, Porto, Portugal, October 3-5, 2009. Proceedings 20}, pages 23--37.
  Springer, 2009.

\bibitem[Donchin et~al.(2000)Donchin, Spencer, and
  Wijesinghe]{donchin2000mental}
Emanuel Donchin, Kevin~M Spencer, and Ranjith Wijesinghe.
\newblock The mental prosthesis: assessing the speed of a p300-based
  brain-computer interface.
\newblock \emph{IEEE transactions on rehabilitation engineering}, 8\penalty0
  (2):\penalty0 174--179, 2000.

\bibitem[Dunn(1961)]{dunn1961multiple}
Olive~Jean Dunn.
\newblock Multiple comparisons among means.
\newblock \emph{Journal of the American statistical association}, 56\penalty0
  (293):\penalty0 52--64, 1961.

\bibitem[Even-Dar et~al.(2002)Even-Dar, Mannor, and Mansour]{even2002pac}
Eyal Even-Dar, Shie Mannor, and Yishay Mansour.
\newblock Pac bounds for multi-armed bandit and markov decision processes.
\newblock In \emph{COLT}, volume~2, pages 255--270. Springer, 2002.

\bibitem[Fan et~al.(2018)Fan, Lewis, and Dauphin]{fan2018hierarchical}
Angela Fan, Mike Lewis, and Yann Dauphin.
\newblock Hierarchical neural story generation.
\newblock \emph{arXiv preprint arXiv:1805.04833}, 2018.

\bibitem[Garivier and Kaufmann(2016)]{garivier2016optimal}
Aur{\'e}lien Garivier and Emilie Kaufmann.
\newblock Optimal best arm identification with fixed confidence.
\newblock In \emph{Conference on Learning Theory}, pages 998--1027. PMLR, 2016.

\bibitem[Hao et~al.(2023)Hao, Jain, Lattimore, Van~Roy, and
  Wen]{hao2023leveraging}
Botao Hao, Rahul Jain, Tor Lattimore, Benjamin Van~Roy, and Zheng Wen.
\newblock Leveraging demonstrations to improve online learning: Quality
  matters.
\newblock \emph{arXiv preprint arXiv:2302.03319}, 2023.

\bibitem[Hershey and Olsen(2007)]{hershey2007approximating}
John~R Hershey and Peder~A Olsen.
\newblock Approximating the kullback leibler divergence between gaussian
  mixture models.
\newblock In \emph{2007 IEEE International Conference on Acoustics, Speech and
  Signal Processing-ICASSP'07}, volume~4, pages IV--317. IEEE, 2007.

\bibitem[Heskebeck et~al.(2022)Heskebeck, Bergeling, and
  Bernhardsson]{heskebeck2022multi}
Frida Heskebeck, Carolina Bergeling, and Bo~Bernhardsson.
\newblock Multi-armed bandits in brain-computer interfaces.
\newblock \emph{Frontiers in Human Neuroscience}, 16, 2022.

\bibitem[Holtzman et~al.()Holtzman, Buys, Du, Forbes, and
  Choi]{holtzmancurious}
Ari Holtzman, Jan Buys, Li~Du, Maxwell Forbes, and Yejin Choi.
\newblock The curious case of neural text degeneration.
\newblock In \emph{International Conference on Learning Representations}.

\bibitem[Jamieson and Nowak(2014)]{jamieson2014best}
Kevin Jamieson and Robert Nowak.
\newblock Best-arm identification algorithms for multi-armed bandits in the
  fixed confidence setting.
\newblock In \emph{2014 48th Annual Conference on Information Sciences and
  Systems (CISS)}, pages 1--6. IEEE, 2014.

\bibitem[Jourdan et~al.(2022)Jourdan, Degenne, Baudry, de~Heide, and
  Kaufmann]{jourdan2022top}
Marc Jourdan, R{\'e}my Degenne, Dorian Baudry, Rianne de~Heide, and Emilie
  Kaufmann.
\newblock Top two algorithms revisited.
\newblock \emph{arXiv preprint arXiv:2206.05979}, 2022.

\bibitem[Jun et~al.(2016)Jun, Jamieson, Nowak, and Zhu]{jun2016top}
Kwang-Sung Jun, Kevin Jamieson, Robert Nowak, and Xiaojin Zhu.
\newblock Top arm identification in multi-armed bandits with batch arm pulls.
\newblock In \emph{Artificial Intelligence and Statistics}, pages 139--148.
  PMLR, 2016.

\bibitem[Kaufmann et~al.(2016)Kaufmann, Capp{\'e}, and
  Garivier]{kaufmann2016complexity}
Emilie Kaufmann, Olivier Capp{\'e}, and Aur{\'e}lien Garivier.
\newblock On the complexity of best arm identification in multi-armed bandit
  models.
\newblock \emph{Journal of Machine Learning Research}, 17:\penalty0 1--42,
  2016.

\bibitem[Krusienski et~al.(2008)Krusienski, Sellers, McFarland, Vaughan, and
  Wolpaw]{krusienski2008toward}
Dean~J Krusienski, Eric~W Sellers, Dennis~J McFarland, Theresa~M Vaughan, and
  Jonathan~R Wolpaw.
\newblock Toward enhanced p300 speller performance.
\newblock \emph{Journal of neuroscience methods}, 167\penalty0 (1):\penalty0
  15--21, 2008.

\bibitem[Kveton et~al.(2021)Kveton, Konobeev, Zaheer, Hsu, Mladenov, Boutilier,
  and Szepesvari]{kveton2021meta}
Branislav Kveton, Mikhail Konobeev, Manzil Zaheer, Chih-wei Hsu, Martin
  Mladenov, Craig Boutilier, and Csaba Szepesvari.
\newblock Meta-thompson sampling.
\newblock In \emph{International Conference on Machine Learning}, pages
  5884--5893. PMLR, 2021.

\bibitem[Ma et~al.(2021)Ma, Huggins, and Kang]{ma2021adaptive}
Tianwen Ma, Jane~E Huggins, and Jian Kang.
\newblock Adaptive sequence-based stimulus selection in an erp-based
  brain-computer interface by thompson sampling in a multi-armed bandit
  problem.
\newblock In \emph{2021 IEEE International Conference on Bioinformatics and
  Biomedicine (BIBM)}, pages 3648--3655. IEEE, 2021.

\bibitem[Ma et~al.(2022)Ma, Li, Huggins, Zhu, and Kang]{ma2022bayesian}
Tianwen Ma, Yang Li, Jane~E Huggins, Ji~Zhu, and Jian Kang.
\newblock Bayesian inferences on neural activity in eeg-based brain-computer
  interface.
\newblock \emph{Journal of the American Statistical Association}, pages 1--12,
  2022.

\bibitem[Manyakov et~al.(2011)Manyakov, Chumerin, Combaz, and
  Van~Hulle]{manyakov2011comparison}
Nikolay~V Manyakov, Nikolay Chumerin, Adrien Combaz, and Marc~M Van~Hulle.
\newblock Comparison of classification methods for p300 brain-computer
  interface on disabled subjects.
\newblock \emph{Computational intelligence and neuroscience}, 2011:\penalty0
  1--12, 2011.

\bibitem[Park and Kim(2012)]{park2012pomdp}
Jaeyoung Park and Kee-Eung Kim.
\newblock A pomdp approach to optimizing p300 speller bci paradigm.
\newblock \emph{IEEE Transactions on Neural Systems and Rehabilitation
  Engineering}, 20\penalty0 (4):\penalty0 584--594, 2012.

\bibitem[Qin and Russo(2022)]{qin2022adaptivity}
Chao Qin and Daniel Russo.
\newblock Adaptivity and confounding in multi-armed bandit experiments.
\newblock \emph{arXiv preprint arXiv:2202.09036}, 2022.

\bibitem[Qin et~al.(2017)Qin, Klabjan, and Russo]{qin2017improving}
Chao Qin, Diego Klabjan, and Daniel Russo.
\newblock Improving the expected improvement algorithm.
\newblock \emph{Advances in Neural Information Processing Systems}, 30, 2017.

\bibitem[Radford et~al.(2019)Radford, Wu, Child, Luan, Amodei, Sutskever,
  et~al.]{radford2019language}
Alec Radford, Jeffrey Wu, Rewon Child, David Luan, Dario Amodei, Ilya
  Sutskever, et~al.
\newblock Language models are unsupervised multitask learners.
\newblock \emph{OpenAI blog}, 1\penalty0 (8):\penalty0 9, 2019.

\bibitem[Russo(2016)]{russo2016simple}
Daniel Russo.
\newblock Simple bayesian algorithms for best arm identification.
\newblock In \emph{Conference on Learning Theory}, pages 1417--1418. PMLR,
  2016.

\bibitem[Russo and Van~Roy(2016)]{russo2016information}
Daniel Russo and Benjamin Van~Roy.
\newblock An information-theoretic analysis of thompson sampling.
\newblock \emph{The Journal of Machine Learning Research}, 17\penalty0
  (1):\penalty0 2442--2471, 2016.

\bibitem[Shang et~al.(2020)Shang, Heide, Menard, Kaufmann, and
  Valko]{shang2020fixed}
Xuedong Shang, Rianne Heide, Pierre Menard, Emilie Kaufmann, and Michal Valko.
\newblock Fixed-confidence guarantees for bayesian best-arm identification.
\newblock In \emph{International Conference on Artificial Intelligence and
  Statistics}, pages 1823--1832. PMLR, 2020.

\bibitem[Simchowitz et~al.(2021)Simchowitz, Tosh, Krishnamurthy, Hsu, Lykouris,
  Dudik, and Schapire]{simchowitz2021bayesian}
Max Simchowitz, Christopher Tosh, Akshay Krishnamurthy, Daniel~J Hsu, Thodoris
  Lykouris, Miro Dudik, and Robert~E Schapire.
\newblock Bayesian decision-making under misspecified priors with applications
  to meta-learning.
\newblock \emph{Advances in Neural Information Processing Systems},
  34:\penalty0 26382--26394, 2021.

\bibitem[Speier et~al.(2011)Speier, Arnold, Lu, Taira, and
  Pouratian]{speier2011natural}
William Speier, Corey Arnold, Jessica Lu, Ricky~K Taira, and Nader Pouratian.
\newblock Natural language processing with dynamic classification improves p300
  speller accuracy and bit rate.
\newblock \emph{Journal of Neural Engineering}, 9\penalty0 (1):\penalty0
  016004, 2011.

\bibitem[Wolpaw et~al.(2018)Wolpaw, Bedlack, Reda, Ringer, Banks, Vaughan,
  Heckman, McCane, Carmack, Winden, et~al.]{wolpaw2018independent}
Jonathan~R Wolpaw, Richard~S Bedlack, Domenic~J Reda, Robert~J Ringer,
  Patricia~G Banks, Theresa~M Vaughan, Susan~M Heckman, Lynn~M McCane,
  Charles~S Carmack, Stefan Winden, et~al.
\newblock Independent home use of a brain-computer interface by people with
  amyotrophic lateral sclerosis.
\newblock \emph{Neurology}, 91\penalty0 (3):\penalty0 e258--e267, 2018.

\end{thebibliography}
}
\newpage

\appendix
\section{Proof of Theorem \ref{thm:fixed_budget}}

We first consider the case when the recommended action is wrong, the subsequent task would use the wrong information. Suppose the prior is as specified in Section \ref{sec:prior_spec}. Define the sub-optimality gap $\Delta_m = \langle A_m^*, \theta_m\rangle-\langle a, \theta_m \rangle$ for any $a\in\cA, a\neq A_m^*$. Note that 
\begin{equation*}
\begin{split}
    \mathbb E\left[\langle A_{m}^*, \theta_m\rangle - \langle \psi_m, \theta_m\rangle\right] &=  \mathbb E\left[\left(\langle A_{m}^*, \theta_m\rangle - \langle \psi_m, \theta_m\rangle\right)\mathbb I\left(\psi_m\neq A_m^*\right)\right]\\
    & \ \ \ + \mathbb E\left[\left(\langle A_{m}^*, \theta_m\rangle - \langle \psi_m, \theta_m\rangle\right)\mathbb I\left(\psi_m= A_m^*\right)\right]\\
    &= \mathbb E\left[\Delta_m\mathbb I\left(\psi_m\neq A_m^*\right)\right]\,,
    \end{split}
\end{equation*}
where $\psi_m$ is the recommeded action at the end of task $m$. Since we assume all the sub-optimal arms have the same mean reward, and the sub-optimality gap is known, the error probability can be bounded through the simple regret:
\begin{equation*}
\mathbb P\left(\psi_m\neq A_m^*\right) = \frac{1}{\Delta_m} \mathbb E\left[\langle A_{m}^*, \theta_m\rangle - \langle \psi_m, \theta_m\rangle\right]\,,
\end{equation*}
where the expectation in the left side is with respect to the prior distribution of $\theta_m$. We employ Proposition 1 of \cite{qin2022adaptivity} for the top-two Thompson sampling for a single task. Define an event $\cE_m=\{\psi_1=A_1^*, \ldots, \psi_{m-1}=A_{m-1}^*\}$. As \cite{qin2022adaptivity} requires the top-two Thompson sampling starting with a correct prior, we decompose the simple regret term based on $\cE_m$ as, 
\begin{equation}\label{eqn:1}
\begin{split}
\mathbb E\left[\langle A_{m}^*, \theta_m\rangle - \langle \psi_m, \theta_m\rangle\right] &=  \underbrace{\mathbb E\left[\left(\langle A_{m}^*, \theta_m\rangle - \langle \psi_m, \theta_m\rangle\right)\mathbb I(\cE_m)\right]}_{I_1} \\
&\ \ \ + \underbrace{\mathbb E\left[\left(\langle A_{m}^*, \theta_m\rangle - \langle \psi_m, \theta_m\rangle\right)\mathbb I(\cE_m^c)\right]}_{I_2}\,.
\end{split}
\end{equation}
Applying Proposition 1 of \cite{qin2022adaptivity}, we obtain that,
\begin{equation}\label{eqn:2}
I_1\leq 6\sqrt{\log(J(1+n))\mathbb H(A_m^*|A_{m-1}^*,\ldots, A_1^*)(1+n)^{-1}}\,,
\end{equation}
where $\mathbb H(\cdot|\cdot)$ is the conditional entropy. To bound $I_2$, we have that,
\begin{equation*}
    \begin{split}
        I_2 &= \mathbb E\left[\left(\langle A_{m}^*, \theta_m\rangle - \langle \psi_m, \theta_m\rangle\right)\mathbb I(\cE_m^c)\right]\\
        &\leq\Delta_m\left(1- \mathbb E\left[\mathbb P\left(\psi_1=A_1^*, \ldots, \psi_{m-1}=A_{m-1}^*\right)\right]\right)\\
        &=\Delta_m\left(1- \mathbb E\left[\prod_{j=1}^{m-1}\mathbb P\left(\psi_j=A_j^*| \psi_{j-1}=A_{j-1}^*,\ldots, \psi_{1}=A_{1}^*\right)\right]\right)\\
        &=\Delta_m\left(1- \mathbb E\left[\prod_{j=1}^{m-1}\left(1-\mathbb P\left(\psi_j\neq A_j^*| \psi_{j-1}=A_{j-1}^*,\ldots, \psi_{1}=A_{1}^*\right)\right)\right]\right)\,.
    \end{split}
\end{equation*}
Applying Proposition 1 of \cite{qin2022adaptivity} again, we have that, 
\begin{equation}\label{eqn:3}
    \mathbb P\left(\psi_j\neq A_j^*| \psi_{j-1}=A_{j-1}^*,\ldots, \psi_{1}=A_{1}^*\right) \leq \frac{6}{\Delta_m}\sqrt{\frac{\log(J(1+n))\mathbb H(A_m^*|A_{m-1}^*,\ldots, A_1^*)}{1+n}}\,.
\end{equation}
Denote $p_m = 6\sqrt{\log(J(1+n))\mathbb H(A_m^*|A_{m-1}^*,\ldots, A_1^*)(1+n)^{-1}}/\Delta_m$. Putting Eqs. \eqref{eqn:1}-\eqref{eqn:3} together, we obtain that,
\begin{equation*}
\begin{split}
\mathbb P\left(\psi_m\neq A_m^*\right) \leq p_m + \left(1-\prod_{j=1}^{m-1}(1-p_j)\right)\,.
\end{split}
\end{equation*}
Summing over all the tasks, we obtain that, 
\begin{equation*}
\sum_{m=1}^M \mathbb P\left(\psi_m\neq A_m^*\right) \leq\sum_{m=1}^M p_m + \sum_{m=1}^M\left(1-\prod_{j=1}^{m-1}(1-p_j)\right)\,,
\end{equation*}
where the second term is the price the agent pays for the wrong prediction.

We next consider the case when the agent is given the identity of the optimal arm $A_m^*$ at the end of task $m$. If the recommended action is wrong, the agent would pay an extra constant price $c$. This is motivated by the BCI setting, i.e., if the system outputs a wrong prediction, the participant would gaze at the backspace and the system would repeat the process until the right word is outputted. In this case, we have that, 
\begin{equation*}
\sum_{m=1}^M \mathbb P\left(\psi_m\neq A_m^*\right) \leq \sum_{m=1}^M p_m\,.
\end{equation*}
This completes the proof.

\section{Verifying the role of prior in the fixed-confidence setting}
\label{sec:allocation}

We study how fast the allocation rule of each arm using \STTS \ converges to the optimal allocation rule and how the information of the prior affects the convergence. Following the prior specified in Section \ref{sec:synthetic} and the computation in \cite{shang2020fixed}, the optimal allocation rule $p^*$ pulls the optimal arm with the probability $\beta=1/2$, and all the other arms with the probability $1/(2J-2)$. We set the prior parameters $\mu=0, \sigma_0^2=1, \Delta=0.5$, the number of arms $J =10$, and we vary $p \in  \cup \{0.1,0.5,0.9\}$. Define $p_{t,a_i} = N_{t,a_i}/t$ as the proportion of selections of arm $a_i$ before round $t$. The algorithm stops when it reaches the pre-specified maximum number of steps $t_{\max} \in \{50,100,150,\ldots,500\}$ per task.

To compare the \STTS \ allocation rule and the optimal allocation rule, we report the KL divergence $\text{KL}(p_{t}, p^*)$ in Figure \ref{fig:allocation}. It is clearly seen that when the prior becomes stronger, the allocation rule induced by \STTS \ converges faster to the optimal allocation rule.

\section{Additional numerical experiments}

\subsection{Additional results of Section \ref{subsec:syn_fix_bud}}
\label{subsec:add_exp1}

\begin{figure}[b!]
\centering
\includegraphics[width=3.3cm]{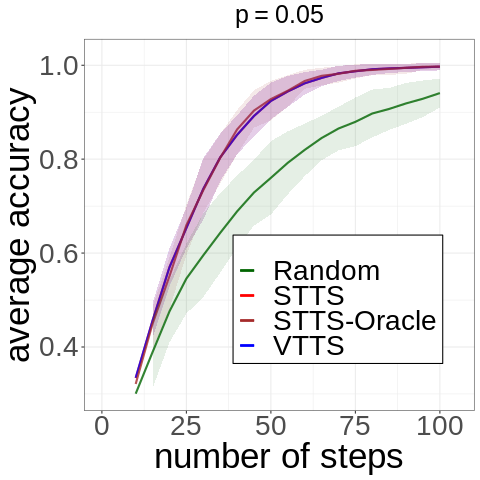}
\includegraphics[width=3.3cm]{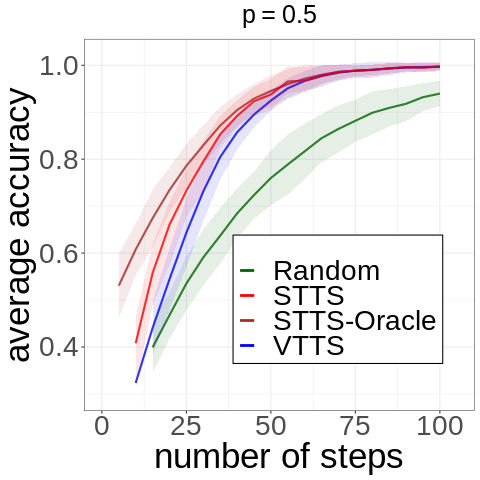}
\includegraphics[width=3.3cm]{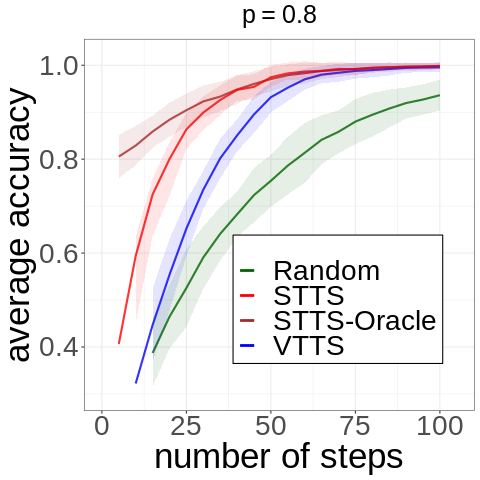}
\includegraphics[width=3.3cm]{SynP300_acc_J=20_p=J_1.png}
\caption{The empirical performance for the fixed-confidence setting with $J=20$. The shaded area represents the 95\% confidence interval.}
\label{fig:FB_synthetic2}
\end{figure}

We report the average accuracy over all $M$ tasks for $J=20$ in Figure \ref{fig:FB_synthetic2}. Similar to the results for $J=10$, \Random \ always performs the worst. When $p=J^{-1}$, indicating a non-informative prior, the performance of \STTS, \texttt{STTS-Oracle}, and \VTTS \ are similar. As $p$ increases and the prior becomes more informative, \texttt{STTS-Oracle} outperforms \STTS, while \STTS outperforms \VTTS significantly. In addition, when $p=1$, \STTS and \texttt{STTS-Oracle} all achieve 100\% accuracy.

\subsection{Additional results of Section \ref{subsec:realdata_res}}
\label{subsec:add_exp2}

\begin{figure}[t!]
\centering
\includegraphics[width=3.4cm]{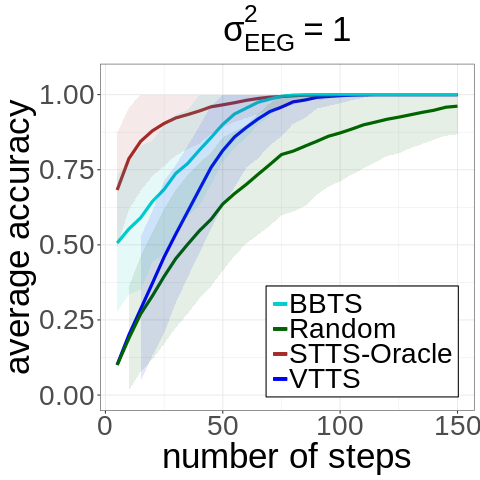}
\includegraphics[width=3.4cm]{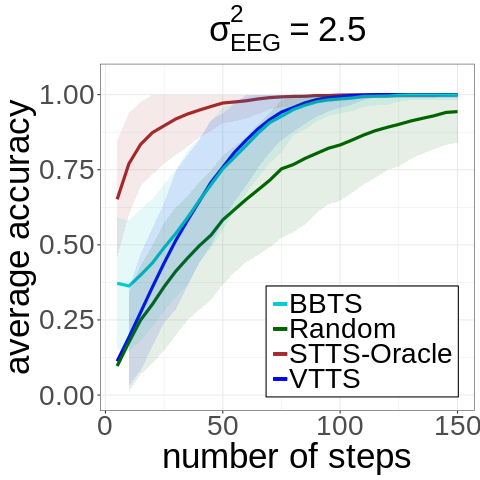}
\includegraphics[width=3.4cm]{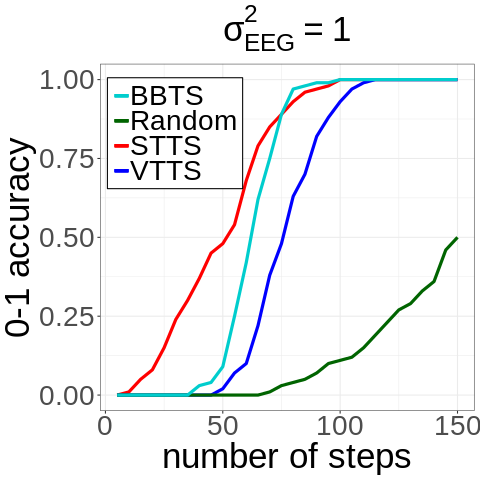}
\includegraphics[width=3.4cm]{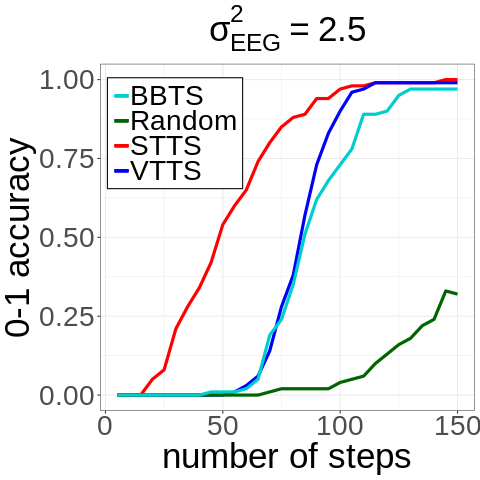}
\caption{The \emph{average accuracy} and \emph{0-1 accuracy} in the fixed budget setting for Prompt 2.}
\label{fig:FB_real_case2} 
\end{figure}

We report the results for the fixed-confidence setting for Prompt 2 in Table \ref{tab:FC_sports} and Figure \ref{fig:FB_real_case2}. The results are similar as those for Prompt 1. By incorporating a language model-informed prior, \STTS \ is able to reduce the required number of stimulus flashes by $25\%$ to $50\%$. Moreover, \STTS \ achieves the highest 0-1 accuracy with a smaller budget than the competing methods.

\begin{table}[t!]
\centering
\begin{tabular}{|c|cc|cc|} 
\hline
Method & $\sigma^2_{\text{EEG}}$  & total steps (std)& $\sigma^2_{\text{EEG}}$  & total steps (std) \\
\hline
\STTS&1&824.80 (171.3)&2.5&1877.4 (258.2)\\
\VTTS&1&1441.1 (221.8)&2.5&2445.8 (252.3)\\
\Random&1&7016.6 (956.8)&2.5&13517. (952.6)\\
\BBTS&1&1421.7 (197.2)&2.5&1694.0 (266.8)\\
\BR&1&2662.0 (140.3)&2.5&4441.7 (271.2)\\
\hline
\end{tabular}
\caption{Total number of steps for Prompt 2.}
\label{tab:FC_sports}
\end{table}

\subsection{Alternative prior specification}
\label{sec:alter_prior}

To show that our proposed algorithm is general and can be coupled with any prior specification, we consider a Gaussian prior specification. The language model $\rho_m$ only requires the previous best arm $A^*_{m-1}$ to define the prior distribution as $\nu_m = \cN(\rho_m(A^*_{m-1}), \sigma_0^2 I)$, where $\rho_m$ represents the LLM and is to be defined later, and $\sigma_0$ is a parameter that determines the standard deviation of the prior distribution. To define $\rho_m$, we first introduce three settings for the matrix $U$: $U^{(1)}$, $U^{(2)}$, and $U^{(3)}$.
\begin{itemize}
    \item For $U^{(1)}$, we set 
    \[
    U^{(1)}_{j_1,j_2}=  
    \begin{cases} & 1, \quad (j_2 - j_1) \in \{1-J,1\}\\
    & 0, \quad \text{otherwise}
    \end{cases},\]
     which means that the optimal arm in the next task is more likely to be the one immediately following the current optimal arm.
    \item For $U^{(2)}$, we set
        \[
    U^{(2)}_{j_1,j_2}=  
    \begin{cases} & 1, \quad (j_1,j_2) \in \{(1,2),(2,3),\ldots,(J-2,1),(J-1,J),(J,J-1)\}\\
    & 0, \quad \text{otherwise}
    \end{cases},\]
    which consists of two groups where the first $(J-2)$ arms are in one group, whereas the last two arms are in the second group.
    \item For $U^{(3)}$,  we set
    \[
    U^{(3)}_{j_1,j_2}=  
    \begin{cases} & 1, \quad (j_2 - j_1) \in \{1-J,1\}\\
    & 0.5, \quad (j_2 - j_1) \in \{2-J,2\}\\
    & 0, \quad \text{otherwise}
    \end{cases},\]
    which possesses more uncertainty than $U^{(1)}$. 
\end{itemize}

Let $e_j$ be the unit vector with length $J$ and the $j$-th entry is 1. We define $\rho_m(A_{m-1}^*) = \mu_0 e_{A_{m-1}^*}^T U$ for $m \ge 2$, where $ \mu_0$ controls the magnitude of the best arm. For $m = 1$, we assign each arm to be the optimal arm with equal probability, and $\nu_1 = N(\mu_0 e_j , \sigma_0^2 I_J)$ if arm $j$ is the optimal arm. The instance $\theta_m$ is sampled from the prior distribution $\nu_m$. 
In this case, we set the posterior mean $\mu_{m,t,j} =  \EE(\theta_{m,j}\in\cdot|\cH_{t,m}\cD_m)$, and the posterior variance $\sigma_{m,t,j}^2 = \Var(\theta_{m,j}\in\cdot|\cH_{t,m}\cD_m)$. Moreover, we set the number of tasks $M=20$, and the agent interacts with $M$ bandit instances with the confidence level $1-\delta = 0.9$, reward noise variance $\sigma^2=1$.

For \VTTS, we specify the non-informative Gaussian prior as $\mu_{0,m,j} = 0$, and $\Sigma_{0,m,j} = 10^2$ for $j \in [J]$. We repeat the experiments 200 times for $J\in \{5,10\}$, $\mu_0\in\{4,5\}$ and $\sigma_0\in\{0.5,1\}$. We report the accuracy and the number of total steps in Table \ref{tab:GaussianPrior_J5} and Table \ref{tab:GaussianPrior_J10}. It is seen from the table that \STTS \ utilizes the smallest number of total steps, and achieves the desired accuracy. When $\mu_0$ is large and $\sigma_0$ small, which means a strong prior information, \STTS \ requires a much smaller number of total steps than \VTTS.  

\begin{table}[t!]
\centering
\begin{tabular}{|cccccccccc|} 
\hline
Method& $\mu_0$& $\sigma_0$&\multicolumn{2}{c}{$U^{(1)}$}&\multicolumn{2}{c}{$U^{(2)}$}&\multicolumn{2}{c}{$U^{(3)}$}&\\
 &&&accuracy & steps&accuracy & steps&accuracy & steps&\\
\hline
\STTS&4&0.5&100.0&60.0&100.0&50.8&98.5&307.5&\\
\VTTS&4&0.5&100.0&253.1&100.0&251.4&99.5&715.9&\\
\Random&4&0.5&100.0&405.1&100.0&406.1&98.0&1351.3&\\
\hline
\STTS&5&0.5&100.0&47.1&100.0&47.3&99.5&76.6&\\
\VTTS&5&0.5&100.0&184.1&100.0&183.3&99.5&370.6&\\
\Random&5&0.5&100.0&295.7&100.0&296.6&99.5&708.5&\\
\hline
\STTS&4&1&98.5&412.1&97.5&429.8&89.5&2008.0&\\
\VTTS&4&1&98.5&644.3&98.5&655.8&89.0&2183.9&\\
\Random&4&1&96.5&1052.1&98.5&1048.1&82.5&3332.3&\\
\hline
\STTS&5&1&99.5&81.8&98.5&80.3&94.5&1115.3&\\
\VTTS&5&1&99.5&267.6&99.5&273.2&93.5&1286.8&\\
\Random&5&1&99.5&449.0&100.0&459.3&94.0&2088.8&\\
\hline
\end{tabular}
\caption{Results of for the Gaussian prior for $J=5$.}
\label{tab:GaussianPrior_J5}
\end{table}

\begin{table}[t!]
\centering
\begin{tabular}{|cccccccccc|} 
\hline
Method& $\mu_0$& $\sigma_0$&\multicolumn{2}{c}{$U^{(1)}$}&\multicolumn{2}{c}{$U^{(2)}$}&\multicolumn{2}{c}{$U^{(3)}$}&\\
 &&&accuracy & steps&accuracy & steps&accuracy & steps&\\
\hline
\STTS&4&0.5&100.0&62.6&98.5&62.2&98.5&455.5&\\
\VTTS&4&0.5&100.0&480.9&100.0&482.1&99.5&1110.1&\\
\Random&4&0.5&100.0&1030.0&99.5&1036.2&98.0&3173.1&\\
\hline
\STTS&5&0.5&100.0&54.9&100.0&54.8&100.0&94.8&\\
\VTTS&5&0.5&100.0&333.1&100.0&336.5&100.0&525.7&\\
\Random&5&0.5&100.0&764.5&100.0&760.8&100.0&1517.3&\\
\hline
\STTS&4&1&98.5&1588.8&96.0&1425.2&93.0&5362.4&\\
\VTTS&4&1&99.0&2162.3&98.0&2145.4&94.5&5700.2&\\
\Random&4&1&95.5&4803.8&95.0&4786.2&88.0&11594.4&\\
\hline
\STTS&5&1&98.5&109.9&100.0&138.0&97.0&2341.1&\\
\VTTS&5&1&100.0&597.4&100.0&616.5&97.5&2771.5&\\
\Random&5&1&100.0&1375.8&98.5&1424.1&96.5&6288.7&\\
\hline
\end{tabular}
\caption{Results of for the Gaussian prior for $J=10$.}
\label{tab:GaussianPrior_J10}
\end{table}

\end{document}